\documentclass[twocolumn,prl,superscriptaddress,nofootinbib]{revtex4-2}

\usepackage{amsmath,amssymb,amsfonts}
\usepackage{graphicx}
\usepackage{xcolor}
\usepackage{hyperref}
\usepackage{braket}

\newcommand{\tr}{\mathrm{tr}}
\newcommand{\Tr}{\operatorname{Tr}}

\begin{document}

\title{Macroscopic Zero-Mode Manifold Isolated by Quantum Chaos}

% \author{A.~Altland}
% \author{C.~Jonay}
% \author{K.~W.~Kim}
% %\email{kunx@cau.ac.kr}
% \author{F.~Pollmann}

\author{C.~Jonay}
\thanks{These authors contributed equally.}
\author{K.~W.~Kim\textsuperscript{*}}
%\email{kunx@cau.ac.kr}
\author{F.~Pollmann}
\author{A.~Altland}

\begin{abstract}
Chaotic many-body spectra are expected to densely fill their energy window. We show that constrained spin chains with chiral symmetry evade this expectation by hosting an exponentially large manifold of symmetry-protected exact zero modes separated from the surrounding spectrum by a sharp gap at zero energy. The gap is generated by chaotic level repulsion, with width set by the number of zero modes times the mean level spacing. We verify this mechanism in an East–West kinetically constrained chain, develop a minimal random-matrix description, and show how the gap can be detected through linear-response spectroscopy.
\end{abstract}

\maketitle

Quantum many-body chaos drives instability and large fluctuations, yet at the same time gives rise to striking forms of universality. Prominent examples are quantum eigenstates homogeneously distributed over effective energy shells as described by the eigenstate thermalization hypothesis \cite{Deutsch1991ETH,Srednicki1994ETH,DAlessio2016FromQuantumChaos} and universal spectral correlations described by random-matrix theory \cite{BohigasGiannoniSchmit1984ChaosRMT,GuhrMuellerGroelingWeidenmueller1998RandomMatrixReview}. Wherever realized, these paradigms establish unifying principles and often connect microscopically distinct classes of quantum systems.

In this work we discuss a somewhat more specialized setting in which the interplay of \textit{symmetry} and many-body chaos gives rise to robust spectral structure. Specifically, we show how a chiral symmetry together with additional unitary symmetries stabilizes exponentially large manifolds of exact many-body zero modes, while chaotic level repulsion in the remaining spectrum expels the surrounding states from the vicinity of zero energy, opening a hard spectral gap. The resulting gap scale is set by the number of zero modes times the mean many-body level spacing.

These conditions are naturally realized in quantum kinetically constrained models, where local degrees of freedom are restricted by the instantaneous state of neighboring sites \cite{RitortSollich2003GlassyKCM,GarrahanLesanovsky2010ThermodynamicsTrajectories,vanHorssenLeviGarrahan2015DynamicsKineticallyConstrained,LanvanHorssenPowellGarrahan2018QuantumSlowRelaxation}. Such systems are known to host symmetry-protected zero modes \cite{SchecterIadecola2018SpectralReflection,Buijsman2022ZeroEnergyPXP,NicolauLjubotinaSerbyn2025FragmentationZeroModes,IvanovMotrunich2025ExactAreaLawScars}, Hilbert-space fragmentation \cite{Sala2020FragmentationDipole,Khemani2020HilbertSpaceShattering}, quantum many-body scars \cite{Bernien2017RydbergScars,Turner2018WeakErgodicityBreakingI,Turner2018WeakErgodicityBreakingII}, and Fock-space cages \cite{JonayPollmann2025FockSpaceCages,TanHuang2025InterferenceCagedScars,BenAmiHeylMoessner2025ManyBodyCages,NicolauLjubotinaSerbyn2025FragmentationZeroModes,MohapatraMoudgalyaBalram2026SpinOneXYCages}. Here we focus on the complementary regime in which the nonzero states remain fully chaotic and ask how an exponentially large zero-mode manifold reshapes the surrounding many-body spectrum.

At the spectral level, the answer can be modeled by chiral random matrix theory. The result is a ``bathtub'' density profile, consisting of a delta-function zero-mode contribution, an empty interval around the symmetry point, and a chaotic bulk outside the gap \cite{Altland2001SpectralTransportBondDisorder}. 
This is the many-body analogue of hard-edge physics in chiral random matrix ensembles \cite{VerbaarschotZahed1993SpectralDensityQCD,Verbaarschot1994SpectrumDiracOperator,
AltlandZirnbauer1997NonstandardSymmetryClasses,
AltlandKimMicklitzRezaeiSonnerVerbaarschot2024QuantumChaosEdge}, with the crucial distinction that the chiral index now grows exponentially with system size, rather than remaining an $\mathcal{O}(1)$ integer.

We demonstrate this mechanism in an East--West constrained spin chain. 
In translation- and inversion-resolved sectors, chiral symmetry enforces $\mu\sim 2^{L/2}$ exact zero modes, while the nonzero spectrum exhibits Gaussian-orthogonal-ensemble correlations after unfolding. 
This coexistence of a protected zero-mode manifold with chaotic bulk dynamics produces a clear dynamical signature that can be resolved through linear-response measurements using appropriately chosen local perturbations.
The remainder of the Letter develops these points quantitatively.

\paragraph{Model.} 
As a simple framework in which our three conditions are jointly
realized, we consider the  next-nearest-neighbor East--West (EW) model, a spin-$1/2$ chain of length $L$
in which domain walls hop left and right, governed by the Hamiltonian
\begin{align}
    H&= \sum_i X_i (P_{i-1}+P_{i+1}) + \beta X_i (P_{i-2}+P_{i+2}),\label{eq:H_EW}
\end{align}
with Pauli operators, $X,Y,Z$ and projectors  $P = (1-Z)/2$ and periodic boundary conditions. This Hamiltonian
anticommutes with $\Gamma = \prod_i Z_i$, realizing the chiral symmetry
$\{H,\Gamma\}=0$, and organizing the hypercubical Hilbert space basis
$\Bbb{Z}_2^L$ into parity disjoint sublattices. In addition, it is unitarily
invariant under lattice translation, $T$, and inversion symmetry, $I$, $[H,T]=
[H,I]=0$.
Despite its strongly constrained dynamics, the Hamiltonian does not exhibit Hilbert space fragmentation--in fact all configurations of the computational basis are connected except the all zero state. 

\paragraph{Macroscopic zero-mode sector.} Within each symmetry sector, chirality
constrains $H$ into block off-diagonal form, $H = \left(\begin{smallmatrix}
0&M\cr M^\dagger & 0 \end{smallmatrix}\right)$, with $N_A\times N_B$ matrices
$M$, implying a number of exact zero modes bounded from below by the sublattice imbalance $|N_A-N_B|$. While this
observation by itself is generic, the crucial question is whether that imbalance
can become parametrically large in a realistic many-body system. This question has a close precedent in the PXP model, where exponentially many zero modes and their distribution over symmetry sectors were counted explicitly~\cite{SchecterIadecola2018SpectralReflection,Buijsman2022ZeroEnergyPXP,Turner2018WeakErgodicityBreakingI,Turner2018WeakErgodicityBreakingII}.
% NOTE: replaced the keys Turner2018WeakErgodicityBreaking and Turner2018QuantumScarredEigenstates by Turner2018WeakErgodicityBreakingI/II, which are the keys used elsewhere -- this removes the duplicate bibliography entries ([9]=[22], [10]=[31]). Please verify the I/II mapping matches the intended papers (Nat. Phys. 14, 745 vs. PRB 98, 155134).
Similarly, in the EW
chain, inversion symmetry leads to an exponential growth with system size. %As shown in the Supplemental Material\fp{Could we give a high level argument here and then refer to the SM for more details?}, 
The combined presence of chirality,
inversion, and translation leads to the bound   
\begin{equation}
    \mu_{\rm tot} \ge 2^{L/2}\label{eq:mu_imbalance},
\end{equation}
for the number of zero modes summed over all sectors. At a high level, this count is set by the number of palindromic basis states, which for even $L$ scales as $2^{L/2}$ (the state of one half of the chain fixes the other by reflection). All of the inversion-protected zero modes reside in the $k=0,\pi$ sectors, the momenta at which inversion and translation commute. Each of these momentum sectors then hosts at least $\mu \equiv \mu_{k=0} \geq 2^{L/2-1}$ zero modes~\cite{Note:mu}, split evenly between the two inversion blocks up to sub-leading corrections (see Supplemental Material for the detailed counting). Numerically, we find that the EW model saturates these bounds, $\mu = 2^{L/2-1}$. The sector-resolved count at $L=18$ has since been independently reproduced in a related rule-based constrained
model~\cite{SilSinghaRoy2026RuleBasedComplexity}. This saturation is not shared by other kinetically constrained models like the PXP.%\footnote{The PXP suffers from Hilbert space fragmentation, while our East--West model is fully connected in the computational basis, except for the all $0$-state.}. 
% NOTE: ED check (beta-independent): dim ker H = 12, 26, 56, 116 for L = 6, 8, 10, 12, matching the sum over sectors of the refined imbalances exactly; k=0 and k=pi host exactly 2^{L/2-1} each (cf. Table 2 of the notes: 20+12 = 32 at L=12).
% The reorganization is extreme: The exponential zero mode degeneracy depletes
% the entire neighborhood of $E=0$, opening a hard gap. The result is a density
% of states with three distinct features: a macroscopic peak at $E=0$, an empty
% region around it, and a smooth bulk for $|E|>\Delta$. This ``bathtub'' profile
% is the spectral fingerprint of an extensive chiral index coexisting with
% ergodic bulk dynamics. The important point is that the gap is not an
% additional assumption, nor a model-specific accident. It is the necessary
% spectral consequence of having an ergodic rectangular chiral block $M$.  
% modes in the $k=0$ sector satisfy $\Gamma I = +1$.

\begin{figure}
\centering
% NOTE: the legend inside panel (b) still reads C mu_sec/D_sec -- please change to C mu/D in the figure source to match the text.
\includegraphics[width=1\columnwidth]{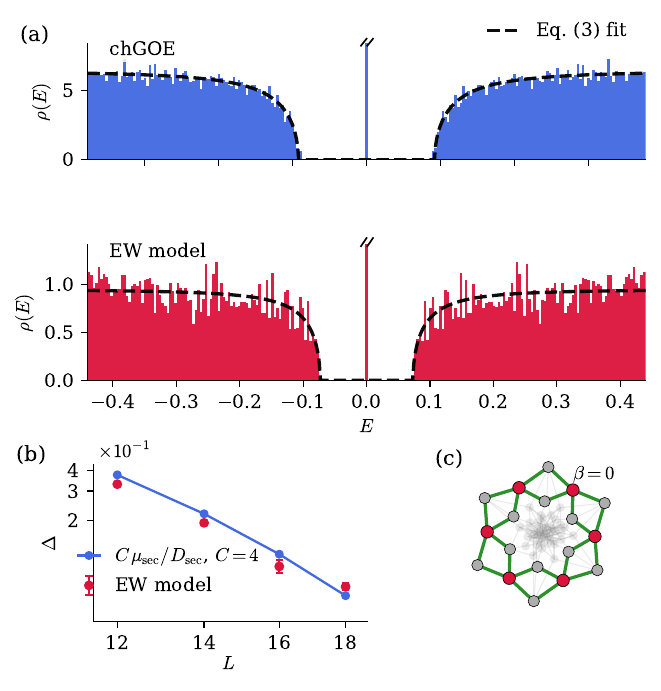}
\caption{Zero-mode manifold and spectral gap.
(a) Density of states near $E=0$ for a  chiral GOE (chGOE) ensemble compared to the East--West model at $L=18$ in the
$k=0$, inversion-even sector ($D=7685$, $\mu_{k=0,I=+}=143$ exact zero modes). The inversion-odd block at
the same momentum contributes another $\mu_{k=0,I=-}=113$, giving $\mu_{k=0}=256=2^{L/2-1}$ in total at
$k=0$. The chGOE comparison uses $N_A=3914$, $N_B=3771$, matching both the
sector dimension ($N_A+N_B=D=7685$) and the chiral imbalance
($N_A-N_B=143$). Eq.~\eqref{eq:MPSpectralDensity} is overlaid as a hard-edge shape
fit: $\mu_{k=0,I=+}$ is fixed by the exact zero-mode count, while the local scale
$\delta$ and amplitude are fitted independently for each
ensemble~\cite{Note:delta}.
(b) Finite-size scaling of the gap $\Delta$; the chGOE scaling $\Delta\sim\mu/D$
captures the EW size dependence up to a model-dependent prefactor.
(c) Representative Fock-space cage in the $\beta=0$ limit.}
\label{fig:dos}
\end{figure}

\paragraph{The gap.} While the  precise number of zero modes $\mu$  solely depends on the organization
of Hilbert space according to the system's unitary symmetries, the additional presence of
chaos leads to a drastic reorganization of the spectrum away from the band
center.  The most apparent consequence of a \emph{large} zero-mode count
is the opening of a  gap around $E=0$ whose width $\Delta\sim \mu \delta$
parametrically exceeds the mean level spacing $\delta$ as we will show below. Intuitively, this gap
can be understood as an extreme manifestation of level repulsion: the typical
spacing between levels in the strongly correlated background of a chaotic system
being given by $\delta$ (up to weak fluctuations), an accumulation of
$\mu$ levels at the symmetry point $E=0$ creates a proportionally larger gap,
$\mu \delta$. 

Phenomenologically, this is captured by chiral random matrix theory (chRMT). In the zero-momentum sector, time-reversal invariance selects the chGOE: the off-diagonal block $M$ is real and Gaussian. Chirality pins $|N_A-N_B|$ states at $E=0$ and pairs the rest into $\pm s_i$, the singular values of $M$. The nonzero eigenvalues of $H^2=(MM^{T})\oplus(M^{T}M)$ are those of the Wishart matrices $MM^{T}$ and $M^{T}M$, distributed by the Marchenko--Pastur law~\cite{MarchenkoPastur1967DistributionEigenvalues}; near $E=0$ this reproduces the universal hard edge of the chiral ensembles %the chiral-class analogue of the Wigner semicircle \cj{The MP distribution is also that of Haar random states, which have nothing to do with chirality, so this sounds a bit too exclusive}
~\cite{VerbaarschotZahed1993SpectralDensityQCD,Verbaarschot1994SpectrumDiracOperator,AltlandZirnbauer1997NonstandardSymmetryClasses}. Its edges $s_\pm=\sigma\bigl(\sqrt{N_A}\pm\sqrt{N_B}\bigr)$ set the gap and the bandwidth: the hard inner edge fixes $\Delta\equiv s_1=s_-\simeq\sigma|N_A-N_B|/2\sqrt{N_A}$, linear in the imbalance (for a single inversion block at $k=0$, $|N_A-N_B|\simeq \mu/2 = 2^{L/2-2}$ as counted above; meanwhile $D = N_A + N_B \simeq 2^L/2L$, the factor $L$
resolving translation and the factor $2$ inversion, so that
$N_A \simeq N_B \simeq 2^L/4L$), while the outer edge gives $W=2s_+$.
% NOTE: sigma is currently undefined in the main text; consider adding "with $\sigma$ the matrix-element scale" after the s_pm equation.
The density of states then takes the universal bathtub form
\begin{equation}
  \rho(E)\simeq \mu\,\delta(E)
  +\frac{1}{\delta}\sqrt{1-\Bigl(\frac{s_1}{E}\Bigr)^{2}}\;
  \Theta\!\bigl(|E|-s_1\bigr),
  \label{eq:MPSpectralDensity}
\end{equation}
where $\delta$ is the local mean spacing of the chaotic bulk, defined by the
plateau value $\rho(E)\simeq 1/\delta$ for $s_1\ll E\ll W$, and $s_1$ the typical value of the first nonzero level. The square-root onset at
$|E|=s_1$ is the chiral-RMT hard edge, symptomatic of an underlying
quantum-critical point. Equivalently $\Delta=s_1\simeq\mu\delta/\pi$ (up to the choice of
spacing convention). The gap is thus simultaneously huge ($\gg\delta$) and microscopic
($\ll W$):
\begin{equation}
  \frac{\Delta}{\delta}\sim\mu\sim 2^{L/2}\to\infty,\qquad
  \frac{\Delta}{W}\sim\frac{\mu}{D}\sim L\, 2^{-L/2}\to 0,
  \label{eq:huge_microscopic}
\end{equation}
with $W=2s_+$ the bulk bandwidth and $D=N_A+N_B$. 
This is the defining feature of the EW model, and the precise point of contrast with the star-graph model (Supplemental Material), where $\mu/D$ stays $O(1)$ and the gap is a finite fraction of $W$.
 
A main result of our present analysis is that individual momentum sectors of the EW model share this behavior. Fig.~\ref{fig:dos}(a) compares the spectral density of the
EW model in the $k=0$ sector to that of the
chGOE. Consistent with the latter, we observe a hard gap
whose width is proportional to $\mu$ and the results found for the microscopic model are in good agreement with RMT predictions as shown in Fig.~\ref{fig:dos}(b). Given the realization of the EW system as a clean model of many-body quantum chaos,  there is some fine print in which we expect deviations from the chGOE prediction. Specifically, the absence of an extensive number of statistically independent model parameters implies a gradual softening of the edge, expected to be of $\mathcal{O}( L \delta)\ll
\mathcal{O}(\mu \delta )$. While this principle may be the reason for a slightly smaller gap size compared to the RMT model, we did not engage in a full exploration of this effect; overall,  the level of agreement between the band-center spectral densities  of the two models appears very good.%, underpinning the presence of fully exposed quantum chaos in the bulk of the EW spectrum.

\begin{figure}
\centering
\includegraphics[width=1\columnwidth]{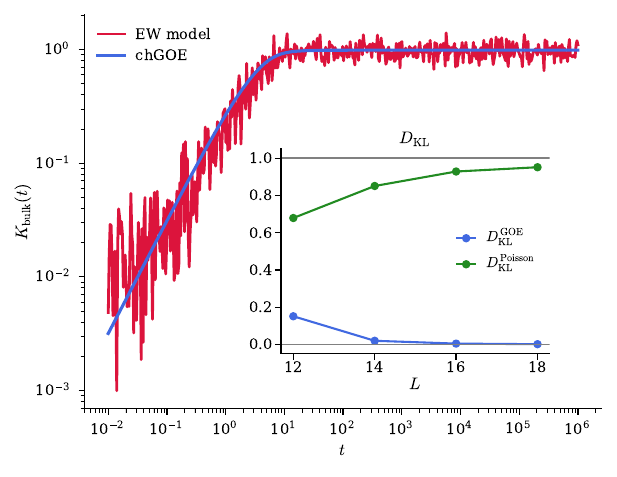}
\caption{Bulk chaos. Unfolded spectral form factor
$K_{\text{bulk}}(t)$ of the EW model for $L=18$ in the $k=0$,
inversion-even sector, versus the chGOE prediction (blue) obtained from the convolution
$\int \rho(x)\, K_{\text{chGOE}}(t/\rho(x))\, dx$, which accounts for the rugged density of states of the EW model, see \cite{GarciaGarcia2025AnatomyScramblingDecoherence}. The GOE ramp emerges beyond a non-parametric Thouless time $t_{\text{Th}} \sim \mathcal{O}(1)$ and saturates onto the Heisenberg plateau. Inset: Kullback--Leibler divergence of the EW level-spacing ratio distribution from GOE (blue) and Poisson (green) references vs.\ $L$; $D_{\text{KL}}^{\text{GOE}} \to 0$ confirms Wigner--Dyson bulk statistics.}
\label{fig:sff}
\end{figure}

\paragraph{Bulk chaos.} While the profile of spectral density
 Eq.~\eqref{eq:MPSpectralDensity} provides evidence for chaotic correlations in
 the system, we here consider its spectral form factor~\cite{Berry1985SemiclassicalSpectralRigidity,Cotler2017BlackHolesRandomMatrices}, $K(t) = |\Tr\,
 e^{-iHt}|^2$, as a more stringent diagnostic. Splitting the trace into a sum
 over zero and finite energy states, the form factor decomposes as
\begin{equation}
    K(t) = \mu^2 + 2\mu \!\sum_{E_n \neq 0}\! \cos(E_n t) + K_{\rm bulk}(t)\,
    \label{eq:sff_decomp}
\end{equation}
where the  constant $\mu^2$ is the contribution from zero modes alone. In the
cross-term the sum over many incommensurate energies  averages
to an approximate zero. Note that this statement can be made exact by introducing an
additional averaging parameter. Turning to the remaining 
contribution $K_\textrm{bulk}$, we expect it to describe bulk spectral
correlations matching those of the GOE (for energies  $\gg \delta$ far
detached from the spectral symmetry point $E=0$, the spectral statistics of the
chGOE reduces to that of the GOE, i.e. the Gaussian ensemble of real symmetric
matrices). However, before making this comparison quantitative,  we need to
remove the influence of the system's inhomogeneous average
density of states through an ``unfolding'' procedure, i.e. a local rescaling to
uniform average density.
%\footnote{Equivalently, the reference random matrix form factor with
%the effect of a non-uniform density of states factored in is given by $\int \rho(x)\, K_{\rm
%GUE}(t/\rho(x))\, dx$
%\cite{GarciaGarcia2025AnatomyScramblingDecoherence}}. 
Figure~\ref{fig:sff} demonstrates that the form factor
rescaled in this manner indeed approaches that of the Gaussian Orthogonal
Ensemble (GOE)  limit for time scales  $\gtrsim t_\textrm{Th}$
exceeding
an effective Thouless time of order $\mathcal{O}(1)$ marking the crossover to a
non-ergodic short time regime.
Focusing on the
complementary limit of long times or  small energy differences, the inset shows the
Kullback-Leibler divergence between the nearest neighbor energy spacing ratio
distribution and those of the GOE and a Poisson ensemble, respectively. 
 The data
indicates convergence to the random matrix distribution in the limit of large
system size. 

%\paragraph{Eigenstates} 

%  \textcolor{red}{Statements:
%  \begin{itemize}
%     \item Cages
%     \item Show IPR
%     \item Page-like entanglement entropy
%     \item Main message: Can engineer cages or all-0 state that is easy to prepare experimentally and relevant for response theory. Rest of spectrum is ergodic. 
%     \item Distribution of IPR across zero energy manifold. 
% \end{itemize}}

% \begin{figure}
% \centering
% \includegraphics[width=1.0\columnwidth]{figures_paper/response.pdf}
% \caption{left: The response principle --- an initial state $\rho$ prepared
% in the $\epsilon=0$ scatters off a probe operator $K$ (grey dot) to either another $E=0$
% state (top), or a bulk state (indicated in green, bottom), where the vertical
% offset indicates the frequency shift $\Omega$ due to the oscillatory
% time-dependence of $K$. Application of the likewise oscillatory observable
% operator $J$ (orange dot) defines the final stae. (The full susceptibility
% kernel $\chi_1$ accounts for a complementary process in which the probe operator
% is applied on the backward propagating amplitude indicated in the bottom).
% Center: exemplary process contributing to the non-linear DC response. Two
% applications of the probe operator transiently shift the frequency carried by
% the intermediate state, which may be either an $E=0$ state, top, or a bulk state
% bottom. Right: Double excitation to an intermediate state of frequency $2\Omega$
% followed by de-excitation via a $2\Omega$-oscillatory obserble.     
% }
% \label{fig:response}
% \end{figure}

\paragraph{Probing the gap.}
Having discussed the principal signatures of the spectral gap, we now
suggest a protocol for its experimental detection. The basic idea is a
combination of  linear response spectroscopy --- a powerful
 technique for probing spectral structures in condensed matter systems --- with the resources provided by
quantum simulators.  Consider our system initialized in some zero energy state,
$\hat\rho(0)=|0\rangle \langle 0|$ (unlike with generic many-body systems, various constrained
systems afford the preparation of select eigenstates, such as $\ket{0}=\ket{00000\dots }$ for the EW
 model).   Assuming its Hamiltonian to be perturbed as $H(t) =
H + V(t)$,
where $V(t) = A(t)K$ is realized as a product of
a probe operator $K$ with an oscillatory modulation
$A(t)\propto \cos(\Omega t+\varphi)$, we are interested in  expectation values
$J(t) \equiv \tr(\hat\rho(t)J)$ of an observation operator $J$. Linear
response analyzes 
 this signal to leading order in $A$, as a signal $J(\Omega)= \chi(\Omega)
V(\Omega)$, observed at a frequency identical to that of the drive, $\Omega$. 
 
 Referring for the first-principles derivation  of the dynamical susceptibility
 $\chi$ to the Supplemental Material, we note that it describes a competition
 of two ``channels'' in the perturbed time evolution of the system. In linear
 response, this evolution comprises  scattering  of the initial state $\braket{0|K|\textrm{int}(t)}$
  to an intermediate state, which
 may either be  a zero energy state $\ket{\textrm{int}(t)}=e^{-i \Omega t}\ket{n}$, or a bulk state $
 \ket{\textrm{int}(t)}=e^{i(E_a-\Omega)t }\ket{a}$, and back into the initial state via $\braket{\textrm{int}(t)|J|0}$. The
 relative weight of the two transient propagation channels is determined by a
 tradeoff between dephasing due to  oscillatory detuning phases, $\sim (E_a-\Omega)$,  and the
 degeneracy of the  $E=0$ state manifold providing a large number of
 intermediate scattering states.  
 
\begin{figure}
\centering
\includegraphics[width=1\columnwidth]{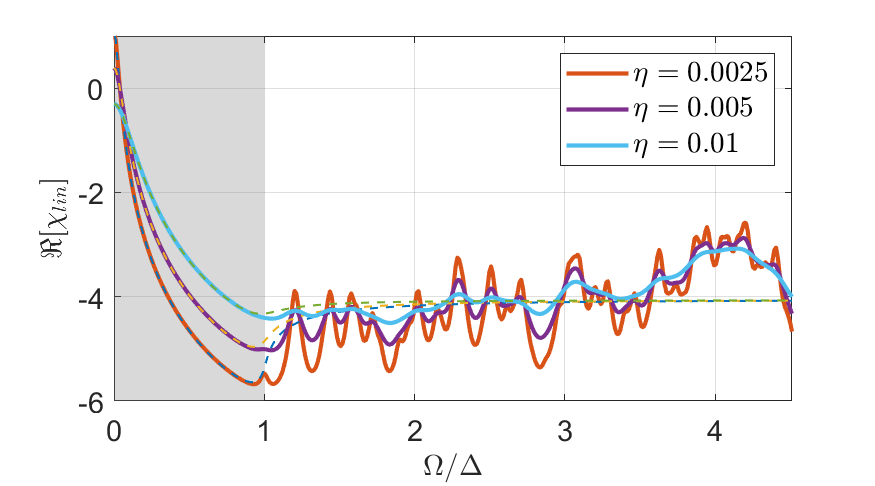}
%{figures_paper/L15.2L18option14.3beta1.4142Nd1W0WR6.4Nw384.Re.lin.12.png}
\includegraphics[width=1\columnwidth]{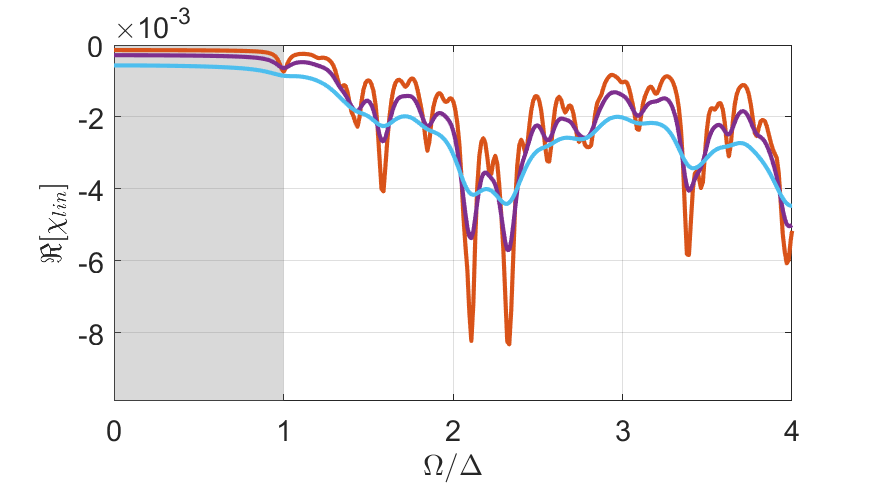}
%{figures_paper/L15.2L18option14.5beta1.4142Nd1W0WR6.4Nw384.Re.lin.12.png}
\caption{
%\fp{I think it would be good to add details about the parameters used like reference to $H$, and provide L,\beta,\dots$}
Linear response for the East--West model~Eq.~\eqref{eq:H_EW} ($L=18, \beta=\sqrt 2$)  under driving with the nearest neighbor spin operator
$K=\sum_i X_{i}X_{{i+1}}\equiv X X$ and observation of $J=-Y X$
(top), vs.  $(K,J)=(XZ,YZ)$ (bottom). 
}
\label{fig:LinearResponse}
\end{figure}

Alongside this competition, the symmetries of $K$ and $J$ relative to the
chiral operator $\Gamma$ are another factor relevant to the frequency
dependence of the signal: if either of the two anticommutes with $\Gamma$,
the zero-energy channel $\ket{\textrm{int}}\propto\ket{n}$ is blocked, as
follows from $\braket{0|X|n}=\braket{0|\Gamma X\Gamma|n}=-\braket{0|X|n}$, for $X=J$ or $K$. The combined effect of these two principles is visible in the linear response data shown in Fig.~\ref{fig:LinearResponse}, where the individual curves differ in the value of a level broadening parameter $\eta$, introduced to mimic the effect of a finite observation time $t\sim \eta^{-1}$ in experiments. For the system size $L=18$ underlying the simulation the level spacing above the gap and the gap itself are given by $\delta =0.0034$ and $\Delta=0.089$, respectively. The level broadening parameters corresponding to the figures are thus $\eta/\delta=0.74, 1.5, 2.9$.     
 
 The top panel shows the real part of the susceptibility, for
 drive operator $K=\sum_i X_{i}X_{{i+1}}\equiv X X$, and observation
 $J=-Y X$, both commutative with $\Gamma$. In this case,  the data indicates a competition
 between a huge (logarithmic units) peak centered at zero drive frequency, whose
 magnitude reflects the contribution of the large number of zero energy modes to
 the sum over intermediate scattering states at low frequencies. (Note the dashed
 lines  representing  Lorentzian fits to this resonant peak.) For frequencies
 reaching the gap offset, individually smaller but now resonant contribution of
 bulk levels becomes significant. The bottom panel shows the susceptibility for
 the cases $(K,J)=(X Z,YZ)$,  now anti-commutative with $\Gamma$. As
 discussed above, the zero mode resonance peak is now absent, such that the
 system responds to driving only for frequencies $\Omega\gg \Delta$. In this
 way, the combined observation of the two types of response reveals information
 about both the number of zero-energy states and their separation from the
 bulk.

% \begin{figure*}
% \centering
% \includegraphics[width=2.1\columnwidth]{figures_paper/Figure__.png}
% \caption{(a) Four different channels contributing to the nonlinear response of
% the system. (i) and (ii) involve frequency doubling, (iii) and (iv) are d.c.,
% where (iv) is distinguished by it containing two zero mode manifold summations.
% (b) Time evolution of the expectation value $\langle \hat O_1(t)-\hat O_1(0)
% \rangle$ after switching on $\hat O_2$ of Eq.~\eqref{eq:ProbeOps} at time $t$
% with visible $2\Omega$ oscillations superimposed on a near stationary
% background. (c) Power spectrum of the signal with visible dominant d.c. signal.
% (d) The scaling $ \chi_\textrm{d.c.}/\chi_\textrm{a.c.}$ for different choices
% of probe and driving operators.  
% }
% \label{fig:scheme}
% \end{figure*}

\paragraph{Discussion.} Many-body systems supporting macroscopically degenerate quantum state manifolds
are rare, examples ranging from Landau levels in ultra-clean two-dimensional
electron gases to the nearly flat bands of twisted bilayer graphene. Here we
identified a mechanism that stabilizes exponentially large manifolds of exactly
degenerate many-body zero-energy states. It combines three ingredients:
unitary crystalline symmetries, a chiral antisymmetry, and many-body
quantum chaos. The first two enforce a zero-mode manifold of dimension
$\mu_{\rm tot}\sim2^{L/2}$, while strong level repulsion generated by quantum chaos
expels the remaining many-body spectrum, opening a hard gap of width
$\Delta\sim\mu \delta$ in each symmetry sector. As long as the protecting symmetries remain intact,
both the degeneracy and the gap are exactly preserved.

Realizing and protecting the required symmetries in condensed-matter
implementations of constrained systems may prove challenging. In this respect,
cold-atom quantum simulators appear particularly promising. They offer clean
and highly controllable environments in which mesoscopic systems can be
realized, for which the many-body gap remains experimentally resolvable despite
the exponentially growing Hilbert space. They furthermore allow for the
preparation of engineered zero-energy states together with time-dependent
control and readout protocols. Taken together, these capabilities should make
it possible to observe both the protected zero-mode manifold and the
accompanying hard gap using experimentally realistic protocols.

The same experimental platform also offers the prospect of engineering
non-translationally invariant synthetic systems in which the gap occupies a
finite fraction of the many-body bandwidth, $\mu/D=\mathcal{O}(1)$, even in the
thermodynamic limit (see the Supplemental Material for a concrete proposal).
Such systems may provide a laboratory for exploring symmetry-protected
degenerate quantum matter, coherent dynamics within exponentially large
protected subspaces, and, ultimately, geometric manipulation of protected
many-body quantum states.

\paragraph{Acknowledgements.}
CJ acknowledges support from the European Union’s
HORIZON-CL4-2022-QUANTUM-02-SGA program under the PASQuanS2.1 project (Grant
Agreement No. 101113690). CJ and FP  acknowledge support from the Deutsche
Forschungsgemeinschaft (DFG) through FOR 5522 (Project-ID 499180199). FP acknowledges support from the DFG under Germany's Excellence Strategy EXC-2111-390814868, TRR 360 (project ID 492547816), and the Munich Quantum Valley, which is supported by the Bavarian state government with funds from the Hightech Agenda Bayern Plus. AA acknowledges support from the DFG under Germany’s Excellence Strategy Cluster of Excellence
Matter and Light for Quantum Computing (ML4Q) EXC
2004/1 390534769 and within the CRC network TR 183
(project grant 277101999) as part of projects A03. KWK acknowledges financial support from
the Basic Science Research Program through the National
Research Foundation of Korea (NRF) funded by the Ministry
of Education (No. RS-2025-00521598) and the Korean Government (MSIT) (No. 2020R1A5A1016518). The numerical data and code used to produce the figures are available in the \href{https://doi.org/10.5281/zenodo.21877168}{Zenodo} repository 21877168.

\bibliographystyle{apsrev4-2}
\bibliography{bib}

\clearpage
\begin{center}
{\large\bfseries Supplemental Material for\\[3pt]
``Quantum Chaos with a Macroscopic Zero-Mode Sector''}
\end{center}

\vspace{1em}

\setcounter{equation}{0}
\setcounter{figure}{0}
\renewcommand{\theequation}{S\arabic{equation}}
\renewcommand{\thefigure}{S\arabic{figure}}

\section{Zero mode bound equation (2)}
\label{app:bound}

We prove that the East--West Hamiltonian~[Eq.~(1)] on a periodic chain of even
length $L$ has at least $2^{L/2}$ exact zero modes,
\begin{equation}
\mu_{\rm tot} \;\geq\; 2^{L/2},
\label{eq:bound-appendix}
\end{equation}
establishing Eq.~(2) of the main text. The argument uses only the chiral and inversion symmetries of $H$; translation is not needed for this bound and enters only in Remarks~1 and~2 below.

\paragraph*{Setup.} Let $\mathcal{H} = (\mathbb{C}^{2})^{\otimes L}$ with
computational basis $\{\ket{s}\}_{s \in \{0,1\}^{L}}$. The chiral operator is
$\Gamma = \prod_{i=1}^{L} Z_{i}$, and the spatial inversion operator is
\begin{equation}
I : \ket{s_{1} s_{2} \cdots s_{L}} \;\longmapsto\; \ket{s_{L} \cdots s_{2} s_{1}}.
\end{equation}
Writing $N(s) = \sum_{i} s_{i}$ for the Hamming weight of a bitstring, $\Gamma$
acts diagonally with eigenvalue $p(s) = (-1)^{N(s)}$. The EW Hamiltonian obeys
\begin{equation}
\{H,\Gamma\} = 0, \qquad [H,I] = 0, \qquad [\Gamma,I] = 0.
\end{equation}

\paragraph*{Step 1: chirality alone is not enough.} Since $\Tr\,\Gamma =
\prod_{i}\Tr Z_{i} = 0$, the global chiral imbalance vanishes and chiral
symmetry by itself implies no exact zero modes.

\paragraph*{Step 2: resolve the inversion sectors.} Because $[\Gamma,I] = 0$,
the inversion eigenspaces $V_{\pm} = \ker(I \mp 1)$ are each preserved by
$\Gamma$. Within each $V_{\pm}$, the anticommutation $\{H,\Gamma\}=0$ forces $H$
into block off-diagonal form between its $\Gamma=+1$ and $\Gamma=-1$ subspaces.
The chiral imbalance of $V_{\pm}$ is
\begin{equation}
Q^{(\pm)} \;=\; \Tr\!\left[\tfrac{1 \pm I}{2}\,\Gamma\right]
\;=\; \tfrac{1}{2}\bigl(\Tr\,\Gamma \,\pm\, \Tr(I\Gamma)\bigr)
\;=\; \pm\tfrac{1}{2}\,\Tr(I\Gamma),
\label{eq:Qpm}
\end{equation}
where the last equality uses $\Tr\,\Gamma = 0$. By the standard chiral pairing
argument, the number of exact zero modes of $H$ in $V_{\pm}$ is bounded below by
$|Q^{(\pm)}|$.

\paragraph*{Step 3: evaluate $\Tr(I\Gamma)$.} In the computational basis,
\begin{equation}
\Tr(I\Gamma) \;=\; \sum_{s}\bra{s}I\Gamma\ket{s}
\;=\; \sum_{s \,:\, Is = s} p(s),
\label{eq:trIG}
\end{equation}
a sum over inversion-fixed strings weighted by parity. An inversion-fixed string
satisfies $s_{i} = s_{L+1-i}$ for all $i$. For $L = 2\ell$, this means $s$ is
determined by its first half $(s_{1},\dots,s_{\ell}) \in \{0,1\}^{\ell}$, chosen
freely, so the number of such strings is exactly
\begin{equation}
\#\{\,s : Is = s\,\} \;=\; 2^{\ell} \;=\; 2^{L/2}.
\end{equation}
Each inversion-fixed string has Hamming weight
$N(s) = 2\sum_{i=1}^{\ell}s_{i}$, which is even, so $p(s) = +1$ for every term
in Eq.~\eqref{eq:trIG}. Therefore
\begin{equation}
\Tr(I\Gamma) \;=\; 2^{L/2}.
\label{eq:trIG-value}
\end{equation}

\paragraph*{Conclusion.} Substituting~\eqref{eq:trIG-value} into~\eqref{eq:Qpm},
\begin{equation}
Q^{(\pm)} \;=\; \pm\,2^{L/2 - 1},
\end{equation}
and by chiral pairing the total number of exact zero modes satisfies
\begin{equation}
\mu_{\rm tot} \;\geq\; |Q^{(+)}| + |Q^{(-)}| \;=\; 2^{L/2},
\end{equation}
which is Eq.~\eqref{eq:bound-appendix}.\hfill$\square$

\paragraph*{Remark 1 (role of translation).} The bound~\eqref{eq:bound-appendix}
used only chirality and inversion. Translation refines this count in two ways.
First, it \emph{localizes} the inversion-protected imbalance: the glide trace
$\Tr(T^{n} I \Gamma)$ equals $2^{L/2}$ for $n$ even and vanishes for $n$ odd
(same counting as above, the fixed points for odd $n$ cancel pairwise by
parity). Fourier-projecting onto momentum $k$ via
$\frac{1}{L}\sum_{n}e^{-ikn}$ then yields
\begin{equation}
Q_{k}^{(+)} - Q_{k}^{(-)}
\;=\;
\begin{cases}
2^{L/2-1}, & k \in \{0,\pi\},\\
0, & \text{otherwise},
\end{cases}
\end{equation}
so the inversion-protected imbalance concentrates in the two momentum sectors
($k=0,\pi$) preserved by inversion (inversion maps $k \to -k$, so a momentum
block is invariant under $I$ only when $k \equiv -k \bmod 2\pi$), with
\begin{equation}
Q_{k=0,\pi}^{(\pm)}
\;=\;
\tfrac{1}{2}\bigl(\nu_{k} \pm 2^{L/2-1}\bigr).
\label{eq:Qkpm}
\end{equation}
Here
\begin{equation}
\nu_{k}
\;=\;
\Tr_{\mathcal{H}_k}\Gamma
\;=\;
\frac{1}{L}\sum_{n}e^{-ikn}\,\Tr(T^{n}\Gamma)
\end{equation}
is the residual chiral trace of the momentum sector, with
\begin{equation}
\Tr(T^{n}\Gamma)
=
\begin{cases}
2^{\gcd(n,L)}, & L/\gcd(n,L)\ \text{even},\\
0, & L/\gcd(n,L)\ \text{odd},
\end{cases}
\end{equation}
so that $\nu_{k} = \mathcal{O}(2^{L/2}/L)$; e.g.\ $\nu_{0}=30$ at $L=18$, and
Eq.~\eqref{eq:Qkpm} reproduces
$Q^{(\pm)}_{k=0} = 143,-113$ of Fig.~1 of the main text. Second, chirality
combined with translation alone already protects $|\nu_{k}|$ zero modes in
\emph{every} momentum sector, including generic $k\neq 0,\pi$. Finally,
translation renders the $(k,I=\pm)$ blocks at $k=0,\pi$ irreducible, which is
what allows direct comparison with the chiral RMT prediction~[Eq.~(3) of the
main text] within a single block of well-defined imbalance
$|N_A-N_B|=|Q^{(\pm)}_{k}|$.

\paragraph*{Remark 2 (saturation).} The bound~\eqref{eq:bound-appendix} is an
inequality in two distinct ways, and exact diagonalization resolves both.
First, within every individual symmetry block the kernel of $H$ is found to
\emph{equal} its chiral imbalance---there are no accidental zero modes. In
particular, the $k=0$ and $k=\pi$ sectors each host exactly
$\mu=2^{L/2-1}$ zero modes (verified sector by sector for $L\leq 12$, and at
$k=0,\pi$ up to $L=18$, where $143+113=256$); the only dynamically
disconnected state, the frozen configuration $\ket{0\cdots 0}$, is itself a
zero mode contained in this count. Second, the \emph{total} number of zero modes
strictly exceeds the inversion bound, since the generic momentum sectors
contribute their translation-protected $|\nu_k|$ zeros as well:
e.g.\ $\mu_{\rm tot}=116>2^{L/2}=64$ at $L=12$. In models with extensive
fragmentation, such as PXP, even the block-wise equality can fail, since
additional zero modes may arise within disconnected Krylov subspaces.

\paragraph*{Remark 3 (generality).} The proof uses nothing about the EW
Hamiltonian except the symmetries
$\{H,\Gamma\}=0$, $[H,I]=0$, $[\Gamma,I]=0$, and even $L$ with periodic
boundary conditions. Equation~\eqref{eq:bound-appendix} therefore applies
verbatim to any local spin-$\tfrac{1}{2}$ chain with these symmetries,
identifying a broad universality class of models hosting an exponentially large
chiral zero-mode manifold.

\section{Star-graph model}

Here we introduce a $k$-local model in which the fraction of zero modes remains
constant with system size. This is in contrast to the East--West model, where
the fraction of zero modes tends to zero in the thermodynamic limit. First, let
us take a step back and think about where the zero modes come from. The starting
point is a Hamiltonian with chiral symmetry,
\begin{align}
H
&=
\begin{pmatrix}
0&M\\
M^{\dagger}&0
\end{pmatrix},
\label{eq:chiral-appendix}
\end{align}
where $M$ is an $N_A\times N_B$ matrix. The number of exact zero modes is given
by the sum of $\dim\ker(M)$ and $\dim\ker(M^{\dagger})$, which can be expressed
as
\begin{align}
\mu
=
|N_A-N_B|
+
2\bigl(\min(N_A,N_B)-\operatorname{rank}(M)\bigr).
\label{eq:mu}
\end{align}
We can thus maximize the number of zero modes either by maximizing the
sublattice imbalance $|N_A-N_B|$ (class~1), or by suppressing
$\operatorname{rank}(M)$ below $\min(N_A,N_B)$ (class~2). The first mechanism
gives a topological invariant, and is what gave us exponentially many zeros in
the East--West model, leading to the bound~\eqref{eq:bound-appendix}. The second
measures rank deficiency: how far $\operatorname{rank}(M)$ falls below what the
matrix shape permits.

As a controlled class~2 example with a finite zero-mode fraction in the
thermodynamic limit, we introduce a $k$-local model on $L+1$ qubits: one
ancilla qubit $\tau$ and $L$ data qubits $s_0,s_1,\ldots,s_{L-1}$. The
Hamiltonian is
\begin{align}
H
=
\tau^+ P_0\,A+\tau^- A^{\dagger}P_0,
\label{eq:star}
\end{align}
where $\tau^{\pm}=(\tau_x\pm i\tau_y)/2$,
$P_0=(1-Z_0)/2$ is the projector onto the $s_0=1$ subspace of the first data
qubit, and $A$ is a random sum of $N_{\rm str}$ $k$-local Pauli strings,
\begin{align}
A
=
\sum_{a=1}^{N_{\rm str}}
g_a
\prod_{i\in S_a}\sigma_i^{\alpha_{a,i}},
\qquad
\alpha_{a,i}\in\{x,y,z\},
\qquad
|S_a|\leq k,
\end{align}
where the coefficients $g_a$ are independent complex Gaussian random variables,
normalized as
$g_a\sim\mathcal{N}_{\mathbb{C}}(0,1/N_{\rm str})$.

A basis state reads
$\ket{\tau,s_0,s_1,\cdots,s_{L-1}}$, where $\tau\in\{+,-\}$ labels the
ancilla and $s_i\in\{0,1\}$ the data qubits. The full state lives in a Hilbert
space of dimension $D=N_A+N_B=2^{L+1}$. The ancilla qubit is the ingredient
that builds in the chiral symmetry,
\begin{align}
\Gamma
=
\tau_z\otimes I_{D/2}
=
\begin{pmatrix}
I_{D/2}&0\\
0&-I_{D/2}
\end{pmatrix},
\end{align}
such that $\{H,\Gamma\}=0$. In the eigenbasis of $\Gamma$, $H$ takes the block
form~\eqref{eq:chiral-appendix} with $M=U_H P_0A$, where $U_H$ acts as a
Hadamard gate on the first data qubit $s_0$ and as the identity on the remaining
data qubits.

The nonzero eigenvalues of $H$ are the singular values of $M$ with both signs.
Let $r=\operatorname{rank}(M)$. Since $U_H$ is unitary, $r$ is bounded by the
rank of the projector,
$r\leq\operatorname{rank}(P_0)=D/4$, and for generic $A$, $r=D/4$ with
probability one. Equation~\eqref{eq:mu} with $N_A=N_B=D/2$ then gives
\begin{align}
\mu
=
D-2r
=
\frac{D}{2},
\qquad
\frac{\mu}{D}
=
\frac{1}{2}.
\end{align}
The zero modes are extensive and occupy a finite fraction of the full Hilbert
space. We compare the chGOE reference, the East--West model, and the star-graph
model in Fig.~\ref{fig:3models}.

\begin{figure}
\centering
\includegraphics[width=.45\columnwidth]{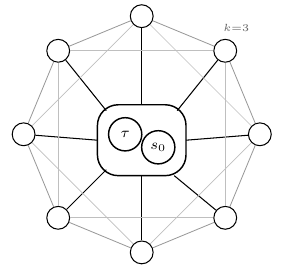}
\includegraphics[width=.45\columnwidth]{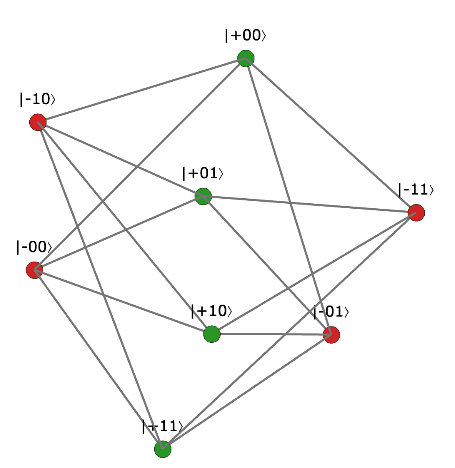}
\caption{Left: Star-graph model connectivity in real space for $L=9$ and
$k=3$. There are two special qubits: the ancilla $\tau$ and the data qubit
$s_0$. For generic $3$-local $A$, the remaining data qubits
$(s_1,\cdots,s_{L-1})$ are connected to their neighbors as shown. Right:
Star-graph model connectivity in Fock space for $L=2$ and $k=3$, red and green
nodes show the two chiral sublattices, labelled by $\tau=\pm$.}
\label{fig:star-graph}
\end{figure}

\begin{figure}
\centering
\includegraphics[width=1\columnwidth]{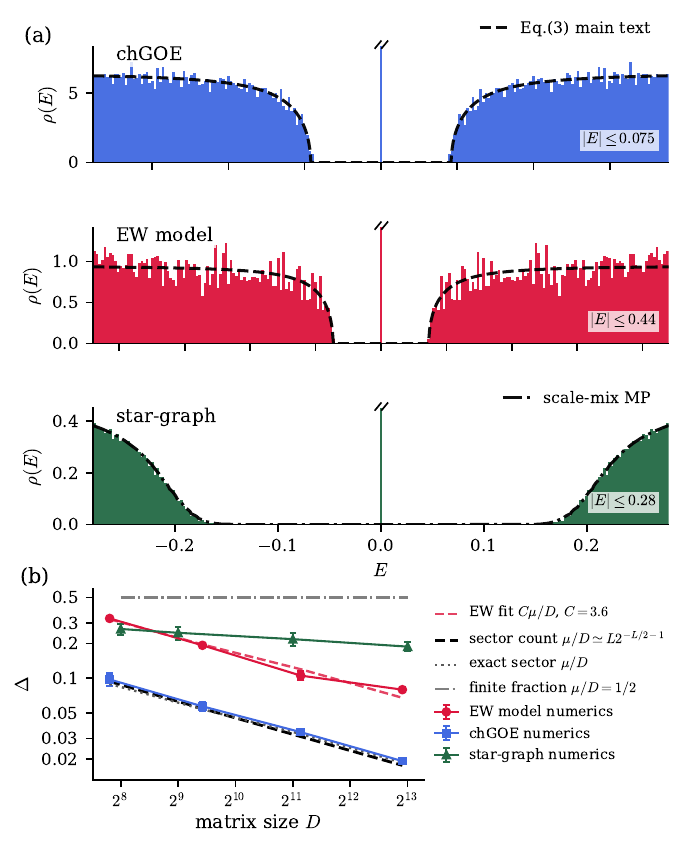}
\caption{Comparison of the chGOE reference, East--West model, and star-graph
model.}
\label{fig:3models}
\end{figure}

\paragraph*{Spectral edge.}
After removing the null rows, the nonzero singular values are those of an
effective rectangular block with $D/4$ rows and $D/2$ columns, hence aspect
ratio
\begin{equation}
q
=
\frac{D/4}{D/2}
=
\frac{1}{2}.
\end{equation}
The left Hadamard $U_H$ is unitary and therefore does not change these singular
values. For a structureless block with overall scale $\sigma$, the
Marchenko--Pastur prediction gives
\begin{equation}
s\in
\bigl[
\sigma(1-\sqrt q),
\,\sigma(1+\sqrt q)
\bigr].
\end{equation}
Thus for $q=1/2$ the nominal lower edge is
\begin{equation}
s_-
=
\sigma\left(1-\frac{1}{\sqrt2}\right)
\simeq
0.293\,\sigma.
\end{equation}
The projection therefore creates two effects at once: an extensive zero-mode
spike, with one half of the full chiral spectrum pinned exactly at $E=0$, and a
repulsion of the remaining nonzero states away from zero. This rectangular
geometry is common to the chGOE reference, the East--West model, and the
star-graph model, so the mere existence of the zero modes and the nominal MP
edge do not by themselves distinguish the three cases.

\paragraph*{Edge shape.}
The distinction is instead in the \emph{shape} of the lower edge. This is
controlled by the variance profile of the projected block,
$\mathbb{E}|M_{xy}|^2$, or equivalently by the weighted row degree
\begin{equation}
w_x
=
\mathbb{E}\,\|M_{x,\cdot}\|^2
=
\sum_y\mathbb{E}|M_{xy}|^2.
\end{equation}
For chGOE this profile is flat. For the translation-invariant East--West model,
the local structure is highly constrained and $w_x$ self-averages, so the edge
is also sharp up to finite-size broadening. In both cases the observed DOS is
well described by the clean-edge form with a square-root onset at the edge.

The star-graph model is different because the projected $k$-local Pauli sum
inherits an inhomogeneous hub-and-spoke connectivity. This produces a
distribution of row weights $w_x$ rather than a single sharply concentrated
value. Weakly connected basis states have a smaller local scale and therefore
contribute singular values below the nominal MP edge. In the present finite-size
data this does not appear as a dramatic deep Lifshitz tail on a linear DOS
scale. Rather, the edge remains broadly MP-like but is visibly rounded. This is
consistent with the Gamma scale-mixture fit in Fig.~\ref{fig:3models}: the
fitted profile-width parameter is relatively large, $\kappa\simeq24$,
corresponding to a moderate relative variance
$\operatorname{Var}(t)=1/\kappa$ in the normalized local scale variable $t$.
Part of the rounding is also attributable to sample-to-sample fluctuations of
the first nonzero eigenvalue $\Delta$. In short, the star-graph model realizes
the same rank-deficiency mechanism as the structureless projected ensemble, but
with a locally inhomogeneous variance profile. The resulting edge is not shifted
by a new counting rule; it is an MP-like edge whose onset is rounded by the
distribution of local Fock-space degrees.

One can generalize the projector $P_0$ to act on a larger number of data qubits,
obtaining an even higher zero-mode fraction.

\section{Linear response}

\begin{figure}
\centering
\includegraphics[width=0.85\columnwidth]{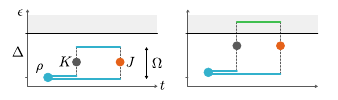}
\caption{Linear-response channels. The initial zero-energy state (blue) is
scattered by the probe operator $K$ (gray dot) into an intermediate state and
returned by the observation operator $J$ (orange dot). The intermediate state
is either another zero mode (left), carrying only the drive-frequency offset
$\Omega$, or a bulk state at energy $E_a$ (right), detuned by $E_a-\Omega$.
The zero-mode channel is enhanced by the $\mu$-fold degeneracy but remains
off-resonant for $\Omega>0$, while the bulk channel becomes resonant once
$\Omega$ exceeds the gap $\Delta$; the competition of the two produces the
structure in Fig.~3 of the main text.}
\label{fig:S3}
\end{figure}

We here provide some details concerning the derivation of the linear-response
formula used in the main text. We consider a general time-dependent
Hamiltonian. Consider the system's time-evolved density matrix,
\begin{equation}
\hat\rho(t)
=
\hat{\mathcal U}_{t,0}\,
\hat\rho\,
\hat{\mathcal U}_{t,0}^{\dagger},
\end{equation}
where
\begin{equation}
\hat{\mathcal U}_{t,0}
=
\mathcal T
\exp\left(
-i\int_0^t\hat H(\tau)\,d\tau
\right)
\end{equation}
is the time-ordered evolution operator. With
$\hat H(t)=\hat H_0+\hat V(t)$ and $\hat V(t)=\hat K A(t)$, the straightforward
expansion in $V$ up to first order yields
\begin{align}
\delta\tilde\rho^{(1)}(t)
&=
-i\int_0^t d\tau\,
\left[
\tilde V(\tau),\hat\rho
\right],
\label{eq:DensityMatrixExpansion}
\end{align}
with the interaction-representation operator
\begin{equation}
\tilde V(t)
=
e^{i\hat H_0t}
\hat V(t)
e^{-i\hat H_0t}
\end{equation}
and
\begin{equation}
\tilde\rho^{(1)}(t)
=
e^{i\hat H_0t}
\hat\rho^{(1)}(t)
e^{-i\hat H_0t}.
\end{equation}
From these terms, we obtain the first-order response contribution to the
physical observable $\hat J$ as
\begin{align}
J(t)
&=
\operatorname{tr}
\left[
\delta\hat\rho^{(1)}(t)\hat J
\right]
=
\int J(\omega)e^{-i\omega t}\,d\omega.
\end{align}
Specifically, the Fourier transform of the response is described by a formal
response kernel whose concrete form is determined in the eigenbasis as follows:
\begin{align}
\chi(\Omega)
&=
-\sum_m
\left[
\frac{J_{nm}K_{mn}}{\Omega-E_m-i\eta}
-
\frac{K_{nm}J_{mn}}{\Omega+E_m-i\eta}
\right]
\nonumber\\
&=
-\sum_m
\frac{
J_{nm}K_{mn}
-
K_{n(\Gamma m)}J_{(\Gamma m)n}
}{
\Omega-E_m-i\eta
},
\label{eq:lin}
\end{align}
where
$J_{nm}=\bra n\hat J\ket m$ and
$J_{(\Gamma m)n}=\bra{\Gamma m}\hat J\ket n$. Here $\ket n$ is the initial
state, which we take to be a zero mode, $E_n=0$. In the second line, $\Gamma$ is
the chiral symmetry operator and
$\hat H_0\Gamma\ket m=-E_m\Gamma\ket m$ is used. The parameter $\eta$ is an
energy uncertainty associated with experiments.

In the main text, we consider operators $J$ and $K$ that either both commute or
both anticommute with $\Gamma$:
\begin{itemize}
\item case 1: $[J,\Gamma]=0$ and $[K,\Gamma]=0$;
\item case 2: $\{J,\Gamma\}=0$ and $\{K,\Gamma\}=0$.
\end{itemize}
Specifically, $(K,J)=(XX,-YX)$ are chosen for case~1, and
$(K,J)=(XZ,YZ)$ for case~2.

The second term in the numerator of Eq.~\eqref{eq:lin} is
\begin{align}
K_{n(\Gamma m)}J_{(\Gamma m)n}
&=
\bra n\Gamma K\ket m
\bra mJ\Gamma\ket n
=
K_{nm}J_{mn},
\end{align}
where in the last equality $\Gamma\ket n=\ket n$ for $E_n=0$ is used. As a
result, only the imaginary part of the matrix element contributes to the
susceptibility:
\begin{align}
\chi(\Omega)
&=
-\sum_m
\frac{
2i\,\Im\!\left[J_{nm}K_{mn}\right]
}{
\Omega-E_m-i\eta
},
\label{eq:lincase1}
\\
&\simeq
C_0\frac{\Omega+i\eta}{\Omega^2+\eta^2}
+
C_1
\int_{\Delta}^{\infty}
dE\,
\rho(E)
\frac{1}{\Omega-E-i\eta}.
\end{align}
In the second line, the susceptibility is divided into contributions from the
zero-energy modes and the nonzero-energy modes; see Fig.~\ref{fig:S3} for a
pictorial description. The coefficients are
\begin{align}
C_0
&=
-\sum_{E_m=0}
2i\,\Im\!\left[J_{nm}K_{mn}\right],
\\
C_1\rho(E)\,dE
&\simeq
-\sum_{E_m\in[E,E+dE]}
2i\,\Im\!\left[J_{nm}K_{mn}\right].
\end{align}

When an operator $K$ anticommutes with $\Gamma$, it does not scatter within the
zero-energy manifold:
\begin{align}
K_{nm}
&=
\bra nK\Gamma^2\ket m
=
-\bra n\Gamma K\Gamma\ket m
=
-K_{nm}
=
0,
\label{eq:scat_zero}
\end{align}
where the fact that a zero mode is its own chiral partner is used:
$\Gamma\ket n=\ket n$ and $\Gamma\ket m=\ket m$ for $E_{n,m}=0$. It follows
that for case~2 the linear response is suppressed as $\Omega\rightarrow0$; see
Fig.~3(bottom) of the main text. On the other hand, for case~1, with $J$ and
$K$ commuting with $\Gamma$, the zero-mode contribution contained in $C_0$
produces the pronounced signal shown in Fig.~3(top) of the main text.

\end{document}